\documentclass{article}

   \usepackage[nonatbib]{neurips_2022}

\usepackage[utf8]{inputenc} 
\usepackage[T1]{fontenc}    
\usepackage{url}            
\usepackage{booktabs}       
\usepackage{amsfonts}       
\usepackage{nicefrac}       
\usepackage{microtype}      
\usepackage{xcolor}         

\title{Towards Better User Requirements: How to Involve Human Participants in XAI Research}

\author{%
   Thu Nguyen \\
   IT University of Copenhagen \\
   Denmark \\
   \texttt{irng@itu.dk} \\
   \And
   Jichen Zhu \\
   IT University of Copenhagen \\
   Denmark \\
   \texttt{jichen.zhu@gmail.com} \\
}

\begin{document}

\maketitle

\begin{abstract}
Human-Center eXplainable AI (HCXAI) literature identifies the need to address user needs. This paper examines how existing XAI research involves human users in designing and developing XAI systems and identifies limitations in current practices, especially regarding how researchers identify user requirements. Finally, we propose several suggestions on how to derive better user requirements. 

\end{abstract}

\section{User Requirements for XAI Explanation}





The research field of eXplainable AI (XAI) has emerged to make AI more transparent and trustworthy to humans by opening the AI black-box and explaining its underlying operation\cite{gunning2017explainable}. While the field has made significant breakthroughs in {\em technical explainability}, it has limited success producing the {\em effective explanations} needed by users\cite{Chromik2021, liao2021human}. As a result, most explanations produced by XAI still lack usability, practical interpretability, and efficacy for real users~\cite{Abdul2018TrendsSystems, Doshi-Velez2017AInterpretability,miller2019explanation,Zhu2018,zhu2021player}. This viewpoint aligns with the one proposed by Liao and Kushney \cite{liao2021human}, where they argue that explanations should address stakeholders' needs. In our paper, we focus on understanding explanation needs from \textit{lay-end users}, who are direct users of the XAI system but have little domain and AI knowledge. This paper's {\em argument} is that a key step towards human-centered XAI and effective explanations is better-defined user requirements through deeper engagement with human users to gather that information. 

User requirements are insights from users about their needs, problems, and the context of use of an interactive system often derived from observing users performing tasks, interviewing users, and conducting focus groups \cite{geis2016cpux}. It is an essential part of the established User-Centered Design (UCD) process for interactive systems in general. Recently there has been growing acknowledgment that user requirements for explanation are necessary to build a usable XAI system \cite{liao2021human}. However, existing work typically attempted to define user requirements based on what the researchers identify as desired qualities of an XAI system \cite{tintarev2007survey, nunes2017systematic, williams2021towards}. While this approach helps to formalize the evaluation criteria, it risks bringing XAI researchers' bias as it is not coming from end users\cite{miller2019explanation}. Therefore, there is a need for methods to extract user requirements from end users directly. 



According to the established UCD process, user requirements should be derived before the XAI system's development to guide the development process towards an outcome that users will find valuable. However, in existing XAI research, only a small portion engages humans in deriving user requirements and other aspects of XAI development. Among them, many diverge from the established UCD design process regarding 1) when user requirements are collected, 2) how to collect user requirements for black-box models, and 3) which user groups are involved. The rest of this section will identify these issues, and the next section proposes approaches to address them.



With a few exceptions (e.g., \cite{ehsan2021expanding, liao2020questioning, putnam2019exploring, Tullio2007}), XAI development teams did not collect user requirements from human users at the early stage of XAI development yet. By contrast, a lot of existing research (e.g., \cite{chen2021towards, dhurandhar2018explanations, Herlocker2000ExplainingRecommendations, rader2018explanations, ribeiro2016should, Stumpf2009}) only engaged human users at the later stage to evaluate the XAI systems. This practice means that a lot of current XAI research is driven by researchers' intuition, and users are only asked at the end to validate the design decisions. 


Among these few projects that obtained user requirements, most seek to understand the user reasoning for Machine Learning models with high interpretability. For example, they focus on how users make sense of Naive Bayes \cite{Wang2019, Stumpf2009} and statistical modeling \cite{Tullio2007}, and use that information to derive user requirements. By contrast, there is a limited understanding of users' sense-making of black box models such as Deep Neural Networks. This is problematic because, without such knowledge, it is difficult to identify effective ways to communicate the information from XAI to users in ways that fit their cognitive needs.




Finally, user requirements gathering in current XAI research typically does not target specific lay-end user groups. As user requirements of explanation highly depend on the context of use \cite{geis2016cpux, liao2020questioning}, the user needs \cite{liao2021human, mohseni2021multidisciplinary}, and user profile \cite{geis2016cpux}. For example, AI developers need explanations to understand the inner working of the model to debug the algorithmic error \cite{liao2021human}. In contrast, non-expert end users will be overwhelmed by the same amount of technical details. For the same user group, whether they use AI in a high-stake or low-stake context will also affect the type of explanation they need. Also, their ability to understand explanations depends on their knowledge of the application domain and AI. Nonetheless, there are many XAI research studies (e.g., \cite{cheng2019explaining, narayanan2018humans, selvaraju2017grad, ribeiro2016should} rely on Amazon Mechanical Turk (AMT) workers, who do not represent a specific domain, needs, or context of use.  We argue that XAI user research should better consider the context of use, domain, and expertise of actual end users.  

\section{Proposed Approaches to Deriving Better HCXAI User Requirements}

 



We first suggest that human-centered XAI research should involve human users at the early stage of development. While there is growing recognition of the importance of the context of use and user needs \cite{liao2021human}, a common practice in the field is to derive user requirements from predefined desiderata such as \cite{lipton2018mythos, Bansal2018}. For example, Lipton~\cite{lipton2018mythos} proposed desiderata such as {\em trust, causality, transferability, informativeness, and fair and ethical decision making}. Bansal proposed {\em actionability} \cite{Bansal2018} as another desiderata for XAI. While these desiderata offer a general direction, they are often too generalized to target the specific requirements in a given context of use. For instance, what is considered trustworthy or informative can differ vastly from AI experts to novice end-users. Conducting user research, especially using qualitative methods (e.g., developing user group profiles, task models, and personas) at the early stage, can be a practical way to supplement the context-free desiderata.



Second, we argue that XAI user requirements should consider how to align users' mental models of the AI (i.e., how the user thinks the AI works) with the system's conceptual model (roughly speaking, how the AI works)\cite{norman1988psychology}. Existing HCI research has established that the usability of a traditional system improves when users' mental models align with the system's conceptual model. Due to the low human interpretability of black-box systems, however, it is unclear whether users, especially non-technical users, are able to construct mental models that completely match the system's conceptual model\cite{Villareale2022}. There are currently no agreements over whether XAI should include all details of system logic\cite{Kulesza2015PrinciplesLearning, Kulesza2013TooModels} or only selective important information\cite{Herlocker2000ExplainingRecommendations, Schaffer2015GettingAnalysis}). Researchers have identified cognitive science theories of how humans make sense of complex information as a potentially fruitful direction to design XAI systems\cite{Wang2019}. We believe that more users studies of how users construct mental models of AI  (e.g., \cite{Villareale2022}) and what cognitive-based design element (e.g., \textit{affordance} \cite{hartson2003cognitive}) can provide much-needed empirical evidence to bridge this knowledge gap. 

Lastly, XAI researchers should identify specific target user groups and develop XAI systems that address their requirements. Any real-world XAI system will be used by specific people (e.g., AI experts, domain experts, non-experts), in particular contexts (work, play, health), and with specific needs. This is especially true because the explanation is social interaction and needs to adapt to the explainees\cite{miller2019explanation}. To improve the usability and efficacy of XAI systems, researchers should identify the above key elements in user requirements. We believe that more research is needed in {\em explanation interfaces}\cite{Chromik2021,mohseni2021multidisciplinary} to study how to effectively {\em communicate} the technical information XAI algorithms generate to users. 

In conclusion, XAI research has yielded significant technical solutions to increase the interpretability and transparency of AI algorithms. To improve its real-world usefulness, we argue that more effort is needed to understand users of XAI systems, especially through better user requirements. 


\section{Acknowledgement}
This work is supported by the Danish Novo Nordisk Foundation Grant NNF20OC0066119.

\small{
\bibliographystyle{unsrt}
\bibliography{ms}}

\end{document}